\documentclass[12pt]{article}

\def\Fint{\rlap{$\Biggl\rfloor$}\Biggl\lceil}

\def\square{\kern1pt\vbox{\hrule height 1.2pt\hbox{\vrule width 1.2pt\hskip 3pt
   \vbox{\vskip 6pt}\hskip 3pt\vrule width 0.6pt}\hrule height 0.6pt}\kern1pt}

\begin{document}

\begin{titlepage}

\begin{flushright}
UFIFT-QG-10-06
\end{flushright}

\vspace{1.5cm}

\begin{center}
{\bf Solving the Effective Field Equations for the Newtonian Potential}
\end{center}

\vspace{.5cm}

\begin{center}
Sohyun Park$^{\dagger}$ and R. P. Woodard$^{\ddagger}$
\end{center}

\vspace{.5cm}

\begin{center}
\it{Department of Physics \\
University of Florida \\
Gainesville, FL 32611}
\end{center}

\vspace{1cm}

\begin{center}
ABSTRACT
\end{center}
Loop corrections to the gravitational potential are usually inferred from
scattering amplitudes, which seems quite different from how the linearized
Einstein equations are solved with a static, point mass to give the
classical potential. In this study we show how the Schwinger-Keldysh
effective field equations can be used to compute loop corrections to
the potential in a way which parallels the classical treatment. We derive
explicit results for the one loop correction from the graviton self-energy
induced by a massless, minimally coupled scalar.

\vspace{.5cm}

\begin{flushleft}
PACS numbers: 04.20.Cv, 02.40.Ky, 04.60.-m, 98.80.-k
\end{flushleft}

\vspace{1.5cm}
\begin{flushleft}
$^{\dagger}$ e-mail: spark@phys.ufl.edu \\
$^{\ddagger}$ e-mail: woodard@phys.ufl.edu
\end{flushleft}
\end{titlepage}

\section{
Introduction}

Everyone who has studied general relativity is familiar with solving
the linearized Einstein equations to derive the first correction due
to a point particle. One expands the full metric about flat space,
\begin{equation}
ds^2 \equiv \Bigl[ \eta_{\mu\nu} + h_{\mu\nu}(t,\vec{x})\Bigr]
dx^{\mu} dx^{\nu} = -c^2 dt^2 + d\vec{x} \cdot d\vec{x} +
h_{\mu\nu}(t,\vec{x}) dx^{\mu} dx^{\nu} \; . \label{invel}
\end{equation}
Linearizing the full Einstein equation gives,
\begin{equation}
R^{\mu\nu} - \frac12 g^{\mu\nu} R = \frac{8\pi G}{c^2} \times T^{\mu\nu}
\; \Longrightarrow \; \mathcal{D}^{\mu\nu\rho\sigma}
h_{\rho \sigma}(t,\vec{x}) = \frac{8\pi G}{c^2} \times M \delta^{\mu}_0
\delta^{\nu}_0 \delta^3(\vec{x}) \; , \label{linEin}
\end{equation}
where the flat space Lichnerowicz operator is,
\begin{equation}
\mathcal{D}^{\mu\nu\rho\sigma} \equiv \frac12 \Bigl[ \eta^{\mu\nu}
\eta^{\rho\sigma} \partial^2 - \partial^{\mu} \partial^{\nu}
\eta^{\rho\sigma} - \eta^{\mu\nu} \partial^{\rho} \partial^{\sigma}
- \eta^{\mu (\rho} \eta^{\sigma) \nu} \partial^2 +
2 \partial^{(\mu} \eta^{\nu) (\rho} \partial^{\sigma)} \Bigr] \; . \label{Lich}
\end{equation}
Up to a linearized gauge transformation, the solution is,
\begin{equation}
h_{00} = \frac{2 G M}{c^2 r} \quad , \quad h_{0i} = h_{i0} = 0 \quad , \quad
h_{ij} = \frac{2 G M}{c^2 r} \, \widehat{r}^i \widehat{r}^j \; , \label{zeroth}
\end{equation}
where $r \equiv \vert \vec{x} \vert$ and $\widehat{r}^i \equiv x^i/r$.

This is simple to understand, and easy to extend to more complicated
sources which can even change with time. Yet we are told that a completely
different technique must be employed in order to include quantum
corrections. Instead of solving the effective field equations, one computes
the gravitational contribution to the amplitude for 2-particle scattering.
Then the inverse scattering problem is used to reconstruct the potentials.
For matter particles one only needs to correct the graviton propagator,
which was done long ago for vectors \cite{AFR,CDH}, spin $\frac12$ fermions
\cite{CD} and scalars \cite{DMC1}. The general result for the long range
Newtonian potential is \cite{MJD,DL},
\begin{equation}
h_{00} \longrightarrow \frac{2 G M}{c^2 r} \Biggl\{ 1 + \Bigl[N_0 +
3 N_{\frac12} + 12 N_1\Bigr] \frac{\hbar G}{45 \pi c^3 r^2} + O(G^2)
\Biggr\} \; ,
\end{equation}
where $N_s$ indicates the number of particles of spin $s$ contributing
to the graviton self-energy, and it should be noted that the $s=0$ result
is for a massless, conformally coupled scalar. (The correction for a 
massless, minimally coupled scalar does not seem to have been published.)
Including gravitons is very much harder and has been the work of decades 
because one must also correct the vertices \cite{AFR,JFD,MK,HL,ABS,KK}. 
The generally accepted result seems to be \cite{gravpot},
\begin{equation}
h_{00} \longrightarrow \frac{2 G M}{c^2 r} \Biggl\{ 1 +
\frac{41 \hbar G}{10 \pi c^3 r^2} + O(G^2) \Biggr\} \; ,
\end{equation}

The scattering amplitude technique is certainly correct, although
there have been many disputes about its application to graviton
corrections \cite{KK}. What bewilders is the contrast between it
and the classical procedure. Why must we do an asymptotic scattering
experiment to infer the gravitational force in quantum field theory
when we can observe orbits over finite times to get it classically?
And what would we do for a dynamic source which was evolving? The
notion of quantum-corrected, effective field equations is more than
half a century old \cite{earlyeff}; why is there not some way of
solving them in analogy to the classical computation?

In fact there is such a technique, and we will employ it here to solve
for the one loop change in the gravitational potentials due to a
massless, minimally coupled scalar field. The results for other massless 
particles have long been known for flat space \cite{MJD,DL}, and the 
method we describe can be used as well in de Sitter background, for 
which the absence of an S-matrix \cite{Witten,Andy} precludes the 
scattering amplitude technique. This paper will serve as a warm-up 
exercise for that more challenging computation, as well as an important 
correspondence limit.

Section 2 is devoted to explaining the general formalism. The actual
computation is done in section 3. Our results are summarized and
discussed in section 4.

\section{Schwinger-Keldysh Effective Field Eqns}

The graviton self-energy $-i [\mbox{}^{\mu\nu} \Sigma^{\rho\sigma}](x;x')$
is the one-particle-irreducible (1PI) 2-point function for the graviton
field $h_{\mu\nu}(t,\vec{x})$. It serves to quantum correct the
linearized Einstein equation (\ref{linEin}),
\begin{equation}
\mathcal{D}^{\mu\nu\rho\sigma} h_{\rho \sigma}(x) + \int \! d^4x'
\Bigl[\mbox{}^{\mu\nu} \Sigma^{\rho\sigma}\Bigr](x;x') \, h_{\rho\sigma}(x')
= \frac{8\pi G}{c^2} M \delta^{\mu}_0 \delta^{\nu}_0 \delta^3(\vec{x})
\; . \label{lineff}
\end{equation}
However, this equation suffers from two embarrassments:
\begin{itemize}
\item{It isn't causal because the in-out self-energy is nonzero for
points $x^{\prime \mu}$ which are spacelike separated from $x^{\mu}$,
or lie to its future; and}
\item{It doesn't produce real potentials $h_{\mu\nu}$ because the
in-out self-energy has an imaginary part.}
\end{itemize}
One can get the right result for a static potential by simply ignoring
the imaginary part \cite{MJD,DL,MK}, but circumventing the limitations
of the in-out formalism becomes more and more difficult as time dependent
sources and higher order corrections are included, and these techniques
break down entirely for the case of cosmology in which there may not even
be asymptotic vacua. It is not that the in-out self-energy is somehow
``wrong''. In fact, it is exactly the right thing to correct the Feynman
propagator for asymptotic scattering computations in flat space. The point 
is rather that equation (\ref{lineff}) doesn't provide the generalization 
we seek of the classical field equation (\ref{linEin}).

The better technique is known as the Schwinger-Keldysh formalism
\cite{earlySK}. It provides a way of computing true expectation
values that is almost as simple as the Feynman diagrams which
produce in-out matrix elements. The Schwinger-Keldysh rules are
best stated in the context of a scalar field $\varphi(x)$ whose
Lagrangian (the space integral of its Lagrangian density) at time
$t$ is $L[\varphi(t)]$. Suppose we are given a Heisenberg state
$\vert \Psi\rangle$ whose wave functional in terms of the operator
eigenkets at time $t_0$ is $\Psi[\varphi(t_0)]$, and we wish to take
the expectation value, in the presence of this state, of a product 
of two functionals of the field operator: $A[\varphi]$, which is
anti-time-ordered, and $B[\varphi]$, which is time-ordered. The
Schwinger-Keldysh functional integral for this is \cite{FW},
\begin{eqnarray}
\lefteqn{\Bigl\langle \Psi \Bigl\vert A[\varphi] B[\varphi] \Bigr\vert \Psi
\Bigr\rangle = \Fint [d\varphi_{\scriptscriptstyle +}]
[d\varphi_{\scriptscriptstyle -}] \,
\delta\Bigl[\varphi_{\scriptscriptstyle -}(t_1) \!-\!
\varphi_{\scriptscriptstyle +}(t_1)\Bigr] } \nonumber \\
& & \hspace{1.9cm} \times A[\varphi_{\scriptscriptstyle -}]
B[\varphi_{\scriptscriptstyle +}]
\Psi^*[\varphi_{\scriptscriptstyle -}(t_0)] e^{i \int_{t_0}^{t_1} dt
\Bigl\{L[\varphi_{\scriptscriptstyle +}(t)] -
L[\varphi_{\scriptscriptstyle -}(t)]\Bigr\}}
\Psi[\varphi_{\scriptscriptstyle +}(t_0)] \; . \qquad \label{fund}
\end{eqnarray}
The time $t_1 > t_0$ is arbitrary as long as it is later than the
latest operator which is contained in either $A[\varphi]$ or $B[\varphi]$.

The Schwinger-Keldysh rules can be read off from its functional representation
(\ref{fund}). Because the same field operator is represented by two different
dummy functional variables, $\varphi_{\scriptscriptstyle \pm}(x)$, the
endpoints of lines carry a ${\scriptscriptstyle \pm}$ polarity. External
lines associated with the anti-time-ordered operator $A[\varphi]$ have
the ${\scriptscriptstyle -}$ polarity whereas those associated with the
time-ordered operator $B[\varphi]$ have the ${\scriptscriptstyle +}$
polarity. Interaction vertices are either all ${\scriptscriptstyle +}$ or
all ${\scriptscriptstyle -}$. Vertices with ${\scriptscriptstyle +}$
polarity are the same as in the usual Feynman rules whereas vertices with
the ${\scriptscriptstyle -}$ polarity have an additional minus sign.
If the state $\vert \Psi\rangle$ is something other than free vacuum then
it contributes additional interaction vertices on the initial value
surface \cite{KOW}.

Propagators can be ${\scriptscriptstyle ++}$, ${\scriptscriptstyle +-}$,
${\scriptscriptstyle -+}$, or ${\scriptscriptstyle --}$. All four polarity
variations can be read off from the fundamental relation (\ref{fund})
when the free Lagrangian is substituted for the full one. It is useful to
denote canonical expectation values in the free theory with a subscript $0$.
With this convention we see that the ${\scriptscriptstyle ++}$ propagator is
just the ordinary Feynman propagator,
\begin{equation}
i\Delta_{\scriptscriptstyle ++}(x;x') = \Bigl\langle \Omega \Bigl\vert
T\Bigl(\varphi(x) \varphi(x') \Bigr) \Bigr\vert \Omega \Bigr\rangle_0 =
i\Delta(x;x') \; , \label{++}
\end{equation}
where $T$ stands for time-ordering and $\overline{T}$ denotes
anti-time-ordering. The other polarity variations are simple to read off and
to relate to the Feynman propagator,
\begin{eqnarray}
i\Delta_{\scriptscriptstyle -+}(x;x') \!\!\! & = & \!\!\! \Bigl\langle \Omega
\Bigl\vert \varphi(x) \varphi(x') \Bigr\vert \Omega \Bigr\rangle_0 \!\!=
\theta(t\!-\!t') i\Delta(x;x') \!+\! \theta(t'\!-\!t) \Bigl[i\Delta(x;x')
\Bigr]^* \! , \qquad \label{-+} \\
i\Delta_{\scriptscriptstyle +-}(x;x') \!\!\! & = & \!\!\! \Bigl\langle \Omega
\Bigl\vert \varphi(x') \varphi(x) \Bigr\vert \Omega \Bigr\rangle_0 \!\!=
\theta(t\!-\!t') \Bigl[i\Delta(x;x')\Bigr]^* \!\!+\! \theta(t'\!-\!t)
i\Delta(x;x') , \qquad \label{+-} \\
i\Delta_{\scriptscriptstyle --}(x;x') \!\!\!\!\! & = & \!\!\! \Bigl\langle
\Omega \Bigl\vert \overline{T}\Bigl(\varphi(x) \varphi(x') \Bigr) \Bigr\vert
\Omega \Bigr\rangle_0 \!\!= \Bigl[i\Delta(x;x')\Bigr]^* . \label{--}
\end{eqnarray}
Therefore we can get the four propagators of the Schwinger-Keldysh formalism
from the Feynman propagator once that is known.

Because external lines can be either ${\scriptscriptstyle +}$ or
${\scriptscriptstyle -}$ in the Schwinger-Keldysh formalism, every
1PI N-point function of the in-out formalism gives rise to $2^N$ 1PI
N-point functions in the Schwinger-Keldysh formalism. For every classical
field $\phi(x)$ of an in-out effective action, the coorresponding
Schwinger-Keldysh effective action must depend upon two fields
--- call them $\phi_{\scriptscriptstyle +}(x)$ and
$\phi_{\scriptscriptstyle -}(x)$ --- in order to access the appropriate
1PI function \cite{lateSK}. For the scalar paradigm we have been considering
this effective action takes the form,
\begin{eqnarray}
\lefteqn{\Gamma[\phi_{\scriptscriptstyle +},\phi_{\scriptscriptstyle -}] =
S[\phi_{\scriptscriptstyle +}] - S[\phi_{\scriptscriptstyle -}]
-\frac12 \int \!\! d^4x \! \int \!\! d^4x' } \nonumber \\
& & \times \left\{\matrix{\!
\phi_{\scriptscriptstyle +}(x) M^2_{\scriptscriptstyle ++}\!(x;x')
\phi_{\scriptscriptstyle +}(x') + \phi_{\scriptscriptstyle +}(x)
M^2_{\scriptscriptstyle +-}\!(x;x') \phi_{\scriptscriptstyle -}(x') \! \cr
\!+ \phi_{\scriptscriptstyle -}(x) M^2_{\scriptscriptstyle -+}\!(x;x')
\phi_{\scriptscriptstyle +}(x') + \phi_{\scriptscriptstyle -}(x)
M^2_{\scriptscriptstyle --}\!(x;x') \phi_{\scriptscriptstyle -}(x') \!}
\right\} + O(\phi^3_{\pm}) , \qquad
\end{eqnarray}
where $S$ is the classical action. The effective field equations are
obtained by varying with respect to $\phi_{\scriptscriptstyle +}$ and then
setting both fields equal \cite{lateSK},
\begin{equation}
\frac{\delta \Gamma[\phi_{\scriptscriptstyle +},\phi_{\scriptscriptstyle -}]
}{\delta \phi_{\scriptscriptstyle +}(x)} \Biggl\vert_{\phi_{\scriptscriptstyle
\pm} = \phi} \!\!\! = \Bigl[\partial^2 - m^2\Bigr] \phi(x)
- \! \int \! d^4x' \Bigl[M^2_{\scriptscriptstyle ++}\!(x;x') +
M^2_{\scriptscriptstyle +-}\!(x;x')\Bigr] \phi(x') + O(\phi^2) . \label{efe}
\end{equation}
The two 1PI 2-point functions we would need to quantum correct the
linearized scalar field equation are $M^2_{\scriptscriptstyle ++}\!(x;x')$
and $M^2_{\scriptscriptstyle +-}\!(x;x')$. Their sum in (\ref{efe}) gives
effective field equations which are causal in the sense that the two 1PI
functions cancel unless $x^{\prime \mu}$ lies on or within the past
light-cone of $x^{\mu}$. Their sum is also real, which neither 1PI function
is separately.

As mentioned before, the point of the present paper is to lay the
groundwork for a computation of the one loop correction to the force
of gravity on de Sitter background. The graviton contribution to the
self-energy was computed some years ago \cite{TW} but has never been
used in the effective field equations. A computation of the scalar
contribution is underway.

Although the current computation will be the first to explore corrections
to a force law, the linearized effective field equations have been studied
on de Sitter background for many simpler models. In scalar quantum
electrodynamics the one loop vacuum polarization was computed and used to
correct for the propagation of dynamical photons \cite{vacpol,PW1}, but not 
yet for the Coulomb force. The one loop scalar self-mass-squared has also 
been used to correct for the propagation of charged, massless, minimally 
coupled scalars \cite{KW}. Both the fermion \cite{fermself} and scalar 
\cite{DW} 1PI 2-point functions of Yukawa theory have been computed and used 
to correct the mode functions. The one and two loop scalar self-mass-squared
of $\lambda \varphi^4$ theory has been computed and used to correct
for the propagation of massless, minimally coupled scalars \cite{scalself}.
In Einstein + Dirac the one loop fermion self-energy has been computed
and used to correct the fermion mode function \cite{Miao}. And the same
thing has been done for scalars in Scalar + Einstein \cite{Kahya}.

\section{Solving for the Potentials}

This section is the heart of the paper. We begin by expressing the
linearized effective field equations in a form which is both manifestly
real and causal. We then explain how these equations can be solved
perturbatively. The hardest step is integrating the one loop
self-energy against the tree order solution. The section closes by
working out the two one loop potentials.

\subsection{Achieving A Manifestly Real and Causal Form}

The basis for our work is a position space result for the one loop
contribution to the 1PI graviton 2-point function from a loop of massless,
minimally coupled scalars, using dimensional regularization and a minimal
choice for the higher derivative counterterms \cite{FW}. (Previous
Schwinger-Keldysh computations of this quantity had been given in momentum
space \cite{Ver} which is not as useful for us.) All four polarization
variations take the form,
\begin{equation}
\Bigl[ \mbox{}^{\mu\nu} \Sigma^{\rho\sigma}_{\scriptscriptstyle \pm \pm}
\Bigr](x;x') = {\rm D}^{\mu\nu\rho\sigma} \Sigma_{\scriptscriptstyle \pm 
\pm}(x;x') \; , \label{gravself}
\end{equation}
where the 4th-order, tensor-differential operator is,
\begin{equation}
{\rm D}^{\mu\nu\rho\sigma} \equiv \Bigl[\eta^{\mu\nu} \partial^2 \!-\!
\partial^{\mu} \partial^{\nu}\Bigr] \Bigl[\eta^{\rho\sigma} \partial^2 \!-\!
\partial^{\rho} \partial^{\sigma}\Bigr] + \frac13 \Bigl[ \eta^{\mu (\rho}
\eta^{\sigma) \nu} \partial^4 \!-\! 2 \partial^{(\mu } \eta^{\nu) (\rho}
\partial^{\sigma)} \!+\! \partial^{\mu} \partial^{\nu} \partial^{\sigma}
\partial^{\rho}\Bigr] \; . \label{loopD}
\end{equation}
The four bi-scalars are,
\begin{equation}
\Sigma_{\scriptscriptstyle \pm\pm}(x;x') = (\pm) (\pm)
\frac{i\kappa^2 \partial^2}{5120 \pi^4} \Biggl[ \frac{\ln( \mu^2
\Delta x^2_{\scriptscriptstyle \pm\pm})}{\Delta x^2_{\scriptscriptstyle
\pm\pm}}\Biggr] \; , \label{scalSig}
\end{equation}
where $\mu^2$ is the usual scale of dimensional regularization, $\kappa^2
\equiv 16 \pi \hbar G/c^3$ is the loop-counting parameter of quantum gravity
and the four Schwinger-Keldysh length functions are,
\begin{eqnarray}
\Delta x^2_{\scriptscriptstyle ++}(x;x') & \equiv & \Bigl\Vert \vec{x} \!-\!
\vec{x}' \Bigr\Vert^2 - c^2 \Bigl( \vert t \!-\! t'\vert \!-\! i \epsilon
\Bigr)^2 \; , \\
\Delta x^2_{\scriptscriptstyle +-}(x;x') & \equiv & \Bigl\Vert \vec{x} \!-\!
\vec{x}' \Bigr\Vert^2 - c^2 \Bigl( t \!-\! t' \!+\! i \epsilon\Bigr)^2 \; , \\
\Delta x^2_{\scriptscriptstyle -+}(x;x') & \equiv & \Bigl\Vert \vec{x} \!-\!
\vec{x}' \Bigr\Vert^2 - c^2 \Bigl( t \!-\! t' \!-\! i \epsilon\Bigr)^2 \; , \\
\Delta x^2_{\scriptscriptstyle --}(x;x') & \equiv & \Bigl\Vert \vec{x} \!-\!
\vec{x}' \Bigr\Vert^2 - c^2 \Bigl( \vert t \!-\! t'\vert \!+\! i \epsilon
\Bigr)^2 \; .
\end{eqnarray}
Although the divergent parts of (\ref{gravself}) have been subtracted off
\cite{FW}, it should be noted that they agree exactly with those originally
found by 't Hooft and Veltman \cite{HV}.

We can achieve a significant simplification by first extracting another
d'Alem\-bert\-ian from (\ref{scalSig}),
\begin{equation}
\Sigma_{\scriptscriptstyle \pm\pm}(x;x') = (\pm) (\pm)
\frac{i\kappa^2 \partial^4}{40960 \pi^4} \Bigl[ \ln^2( \mu^2
\Delta x^2_{\scriptscriptstyle \pm\pm}) - 2 \ln(\mu^2
\Delta x^2_{\scriptscriptstyle \pm\pm}) \Bigr] \; .
\end{equation}
Now define the position and temporal separations, and the associated
invariant length-squared,
\begin{equation}
\Delta r \equiv \Vert \vec{x} \!-\! \vec{x}'\Vert \qquad , \qquad
\Delta t \equiv t \!-\! t' \qquad , \qquad \Delta x^2 \equiv \Delta
r^2 - c^2 \Delta t^2 \; .
\end{equation}
The ${\scriptscriptstyle ++}$ and ${\scriptscriptstyle +-}$
logarithms can be expanded in terms of their real and imaginary
parts,
\begin{eqnarray}
\ln(\mu^2 \Delta x^2_{\scriptscriptstyle ++}) & = & \ln(\mu^2 \vert
\Delta x^2\vert) + i \pi \, \theta(-\Delta x^2) \; , \\
\ln(\mu^2 \Delta x^2_{\scriptscriptstyle +-}) & = & \ln(\mu^2 \vert
\Delta x^2\vert) - i \pi \, {\rm sgn}(\Delta t) \theta(-\Delta x^2)
\; .
\end{eqnarray}
The ${\scriptscriptstyle ++}$ and ${\scriptscriptstyle +-}$
logarithms agree for spacelike separation ($\Delta x^2 > 0$), and
for $t' > t$, whereas they are complex conjugates of one another for
$x^{\prime \mu} = (ct',\vec{x}')$ in the past light-cone of $x^{\mu} =
(ct,\vec{x})$. Hence the sum of $\Sigma_{\scriptscriptstyle
++}(x;x')$ and $\Sigma_{\scriptscriptstyle +-}(x;x')$ is both causal
and real,
\begin{equation}
\Sigma_{\scriptscriptstyle ++}(x;x') + \Sigma_{\scriptscriptstyle
+-}(x;x') = -\frac{\kappa^2 \partial^4}{10240 \pi^3} \,
\theta(c\Delta t \!-\! \Delta r) \Bigl[ \ln(-\mu^2 \Delta x^2) \!-\!
1\Bigr] \; .
\end{equation}

Let us assume that the state is released in free vacuum at time
$t=0$. Our final result for the linearized, one loop effective field
equations is,
\begin{eqnarray}
\lefteqn{\mathcal{D}^{\mu\nu\rho\sigma} h_{\rho\sigma}(t,\vec{x}) -
\frac{\kappa^2 {\rm D}^{\mu\nu\rho\sigma} \partial^4}{10240 \pi^3}
\int_0^{t} \!\! dct' \int \!\! d^3x' \, \theta(c\Delta t \!-\! \Delta
r) } \nonumber \\
& & \hspace{3cm} \times \Bigl[\ln(-\mu^2 \Delta x^2) \!-\! 1\Bigr]
h_{\rho\sigma}(t',\vec{x}') = \frac{8\pi G M}{c^2} \, \delta^{\mu}_0
\delta^{\nu}_0 \delta^3( \vec{x}) \; . \qquad \label{linoneeff}
\end{eqnarray}
Recall that $\kappa^2 \equiv 16 \pi \hbar G/c^3$ is the loop
counting parameter of quantum gravity, the Lichnerowitz operator
$\mathcal{D}^{\mu\nu\rho\sigma}$ was given in (\ref{Lich}) and the
4th order differential operator ${\rm D}^{\mu\nu\rho\sigma}$ was given 
in (\ref{loopD}).

\subsection{Solving the Equation Perturbatively}

There is no point in trying to solve equation (\ref{linoneeff})
exactly because it only includes the one loop graviton self-energy.
A better approach is to seek a perturbative solution in powers of
the loop counting parameter $\kappa^2$,
\begin{equation}
h_{\mu\nu}(t,\vec{x}) = \sum_{\ell = 0}^{\infty} \kappa^{2\ell}
h^{(\ell)}_{\mu\nu}(t,\vec{x}) \; . \label{series}
\end{equation}
Of course the $\ell = 0$ term obeys the linearized Einstein equation
whose solution in Schwarzschild coordinates is,
\begin{equation}
\mathcal{D}^{\mu\nu\rho\sigma} h^{(0)}_{\rho\sigma}(t,\vec{x}) =
\frac{8\pi G M}{c^2} \, \delta^{\mu}_0 \delta^{\nu}_0
\delta^3(\vec{x}) \; \Longrightarrow \; h^{(0)}_{00} = \frac{2 G
M}{c^2 r} \, , \, h^{(0)}_{ij} = \frac{2 G M}{c^2 r} \,
\widehat{r}^i \widehat{r}^j \; . \label{h0eqn}
\end{equation}
The one loop correction $h^{(1)}_{\mu\nu}$ obeys the equation,
\begin{equation}
\mathcal{D}^{\mu\nu\rho\sigma} h^{(1)}_{\rho\sigma}(t,\vec{x}) =
\frac{{\rm D}^{\mu\nu\rho\sigma} \partial^4}{10240 \pi^3} \int_0^t \!\!
dt' \!\! \int\!\! d^3x' \, \theta(\Delta t \!-\! \Delta r) \Bigl[
\ln(-\mu^2 \Delta x^2) \!-\! 1\Bigr]
h^{(0)}_{\rho\sigma}(t',\vec{x}') \; . \label{h1eqn}
\end{equation}
Finding the two loop correction $h^{(2)}_{\mu\nu}$ would require the
two loop self-energy, which we do not have, so $h^{(1)}_{\mu\nu}$ is
as high as we can go.

\subsection{The One Loop Source Term}

In this subsection we evaluate the right hand side of
equation(\ref{h1eqn}), which sources the one loop correction
$h^{(1)}_{\mu\nu}(t,\vec{x})$. This is done in three steps: we first
perform the integral, then act the $\partial^4$, and finally act the
${\rm D}^{\mu\nu\rho\sigma}$.

From the form of the tree order potentials (\ref{h0eqn}) it is
apparent that we need two integrals. The first comes from
$h^{(0)}_{00}$,
\begin{equation}
\int_0^t \!\! dt' \! \int \!\! d^3x' \, \theta(\Delta t \!-\! \Delta
r) \Bigl[ \ln(-\mu^2 \Delta x^2) \!-\! 1\Bigr] \times \frac1{\Vert
\vec{x}'\Vert} \equiv F(t,r) \; .
\end{equation}
The second integral derives from the other nonzero potential
$h^{(0)}_{ij}$. Its trace part is obviously the same as $F(t,r)$,
and we shall call its traceless part $G(t,r)$,
\begin{eqnarray}
\lefteqn{\int_0^t \!\! dt' \! \int \!\! d^3x' \, \theta(\Delta t
\!-\! \Delta r) \Bigl[ \ln(-\mu^2 \Delta x^2) \!-\! 1\Bigr] \times
\frac{\widehat{r}^{\prime i} \widehat{r}^{\prime j}}{\Vert
\vec{x}'\Vert} } \nonumber \\
& & \hspace{3cm} \equiv \frac12 \Bigl[\delta^{ij} \!-\!
\widehat{r}^i \widehat{r}^j\Bigr] F(t,r) - \frac12 \Bigl[\delta^{ij}
\!-\! 3 \widehat{r}^i \widehat{r}^j\Bigr] G(t,r) \; , \qquad \\
& & \hspace{3cm} = \frac13 \delta^{ij} F(t,r) + \frac12 \Bigl[3
\widehat{r}^i \widehat{r}^j \!-\! \delta^{ij} \Bigr] \Bigl[ G(t,r)
\!-\! \frac13 F(t,r)\Bigr] \; . \qquad
\end{eqnarray}
The integrals are tedious but straightforward and give the following
results for $F(t,r)$ and the combination $G(t,r) - \frac13 F(t,r)$,
\begin{eqnarray}
\lefteqn{F(t,r) = \frac{4\pi}{r} \Biggl\{ \frac{r^4}6 \ln(2\mu r) \!-\!
\frac{25}{72} r^4 \!+\! \frac{11}{18} r^3 ct \!-\! \frac{11}{18} r
c^3 t^3 + \Bigl[ \frac1{12} (ct\!+\! r)^4 } \nonumber \\
& & \hspace{.2cm} - \frac{r}6 (r\!+\!ct)^3\Bigr] \ln\Bigl[ \mu (ct
\!+\! r)\Bigr] - \Bigl[\frac1{12} (ct \!-\! r)^4 \!+\! \frac{r}6
(ct \!-\! r)^3\Bigr] \ln\Bigl[\mu (ct \!-\! r)\Bigr] \Biggr\} , \qquad \\
\lefteqn{G(t,r) \!-\! \frac{F(t,r)}3 = \frac{4\pi}{r} \Biggl\{
-\frac{r^4}9 \ln(2\mu r) \!+\! \frac{23}{108} r^4 \!-\!
\frac{199}{675} r^3 ct \!-\! \frac{13}{135} r c^3 t^3 \!+\! \frac{2
c^5 t^5}{45 r} } \nonumber \\
& & \hspace{.2cm} + \Bigl[- \frac{(ct\!+\! r)^6}{45 r^2} \!+\!
\frac{2}{15} \frac{(ct \!+\!r)^5}{r} \!-\! \frac{5}{18} (ct \!+\! r)^4
\!+\! \frac29 r (ct\!+\!r)^3\Bigr] \ln\Bigl[ \mu (ct \!+\! r)\Bigr]\nonumber \\
& & \hspace{.2cm} + \Bigl[\frac{(ct\!-\! r)^6}{45 r^2} \!+\!
\frac{2}{15} \frac{(ct \!-\!r)^5}{r} \!+\! \frac{5}{18} (ct \!-\! r)^4
\!+\! \frac29 r (ct\!-\!r)^3\Bigr] \ln\Bigl[ \mu (ct \!-\! r)\Bigr]
\Biggr\} . \qquad
\end{eqnarray}

The next step is acting the two d'Alembertians. This purges all the
time dependent terms,
\begin{eqnarray}
\lefteqn{ \partial^4 F(t,r) = \frac{4\pi}{r} \times 4 \ln(2 \mu r)
\; , } \\
\lefteqn{ \partial^4 \Biggl\{ \frac13 \delta^{ij} F(t,r) + \frac12
\Bigl[3 \widehat{r}^i \widehat{r}^j - \delta^{ij}\Bigr] \Bigl[G(t,r)
\!-\! \frac13 F(t,r)\Bigr] \Biggr\} } \nonumber \\
& & \hspace{3cm} = \frac{4\pi}{r} \Biggl\{ \frac43 \delta^{ij}
\ln(2\mu r) + \Bigl[3 \widehat{r}^i \widehat{r}^j \!-\!
\delta^{ij}\Bigr] \Bigl[ \frac43 \ln(2\mu r) \!-\! 2\Bigr] \Biggr\}
\; . \qquad
\end{eqnarray}
At this stage the linearized, one loop effective field equations
(\ref{linoneeff}) take the form,
\begin{equation}
\mathcal{D}^{\mu\nu\rho\sigma} h^{(1)}_{\rho\sigma}(t,\vec{x}) =
\frac{G M}{1280 \pi^2 c^2} \, {\rm D}^{\mu\nu\rho\sigma}
f_{\rho\sigma}(\vec{x}) \; ,
\end{equation}
where the nonzero components of the tensor $f_{\rho\sigma}(\vec{x})$
are,
\begin{eqnarray}
f_{00}(\vec{x}) & = & \frac4{r} \, \ln(2 \mu r) \; , \\
f_{ij}(\vec{x}) & = & \delta^{ij} \times \frac43 \frac{\ln(2\mu
r)}{r} + \Bigl[ 3 \widehat{r}^i \widehat{r}^j \!-\!
\delta^{ij}\Bigr] \Bigl[ \frac43 \frac{\ln(2\mu r)}{r} \!-\!
\frac2{r}\Bigr] \; . \qquad
\end{eqnarray}

It remains only to act the operator ${\rm D}^{\mu\nu\rho\sigma}$ on 
$f_{\rho\sigma}(\vec{x})$. The first two derivatives give,
\begin{eqnarray}
\partial^{\rho} \partial^{\sigma} f_{\rho\sigma} = \frac{4}{r^3} & \!\!,\!\! &
\partial_j f_{ij} = \widehat{r}^i \, \frac{[4 \ln(2 \mu r) \!-\! 4]}{r^2}
\; , \qquad \\
\partial^2 f_{00} = -\frac{4}{r^3} & \!\!,\!\! & \partial^2 f_{ij} = 
\delta^{ij} \frac{[8 \ln(2 \mu r) \!-\! 12]}{r^3} - \widehat{r}^i \widehat{r}^j 
\frac{[24 \ln(2 \mu r) \!-\! 32]}{r^3} \; , \qquad \\
\partial_i \partial_j f_{00} = \partial_k \partial_i f_{jk} & \!\!,\!\! & 
\partial_i \partial_k f_{jk} = \delta^{ij} \, \frac{[4 \ln(2\mu r)\!-\!4]}{r^3} 
- \widehat{r}^i \widehat{r}^j \, \frac{[12 \ln(2\mu r) \!-\! 16]}{r^3} 
\; . \qquad
\end{eqnarray}
The source term can then be expressed in terms of two more derivatives
of the quantities $g_{\mu\nu}(\vec{x}) \equiv \nabla^2 f_{\mu\nu}(\vec{x})$
and $g(\vec{x}) \equiv \partial_i \partial_j f_{ij}(\vec{x})$,
\begin{equation}
\mathcal{D}^{\mu\nu\rho\sigma} h^{(1)}_{\rho\sigma}(t,\vec{x}) =
\frac{G M}{1280 \pi^2 c^2} \Bigl\{-\eta^{\mu\nu}
\nabla^2 g + \frac43 \partial^{\mu} \partial^{\nu} g + \frac13
\nabla^2 g^{\mu\nu} - \frac23 \partial_{\rho} \partial^{(\mu} g^{\nu) \rho}
\Bigr\} \; .
\end{equation}
The final reduction employs the identities,
\begin{eqnarray}
\nabla^2 g = \frac{24}{r^5} & , & \partial_i \partial_j g = -\frac{12}{r^5} \,
\delta^{ij} \!+\! \frac{60}{r^5} \, \widehat{r}^i \widehat{r}^j \; , \\
\nabla^2 g_{00} = -\frac{24}{r^5} & , & \nabla^2 g_{ij} = -\frac{48}{r^5} \,
\delta^{ij} \!+\! \frac{120}{r^5} \, \widehat{r}^i \widehat{r}^j \; , \\
\partial_k g_{jk} = -\frac{12}{r^4} \, \widehat{r}^j & , & \partial_i
\partial_k g_{jk} = -\frac{12}{r^5} \, \delta^{ij} \!+\! \frac{60}{r^5}
\, \widehat{r}^i \widehat{r}^j \; .
\end{eqnarray}
The nontrivial components of the effective field equations are,
\begin{eqnarray}
\mathcal{D}^{0 0 \rho\sigma} h^{(1)}_{\rho\sigma}(t,\vec{x}) & = &
\frac{G M}{80 \pi^2 c^2} \times \frac1{r^5} \; , \label{time} \\
\mathcal{D}^{i j \rho\sigma} h^{(1)}_{\rho\sigma}(t,\vec{x}) & = &
\frac{G M}{80 \pi^2 c^2} \times \Bigl\{-\frac{3 \delta^{ij}}{r^5} 
\!+\! \frac{5 \widehat{r}^i \widehat{r}^j}{r^5} \Bigr\} \; . \label{space}
\end{eqnarray}

\subsection{The One Loop Potentials}

We wish to express the one loop potentials in Schwarzschild coordinates
so their nonzero components take the form,
\begin{equation}
h^{(1)}_{00}(\vec{x}) = a(r) \qquad , \qquad h^{(1)}_{ij}(\vec{x}) = 
\widehat{r}^i \widehat{r}^j b(r) \; .
\end{equation}
Acting the Lichnerowitz operator (\ref{Lich}) on these gives,
\begin{eqnarray}
\mathcal{D}^{00\rho\sigma} h^{(1)}_{\rho\sigma} & = & \frac{b'}{r} + 
\frac{b}{r^2} \; , \label{00comp} \\
\mathcal{D}^{ij\rho\sigma} h^{(1)}_{\rho\sigma} & = & \delta^{ij}
\Bigl[- \frac{a''}{2} - \frac{a'}{2 r} - \frac{b'}{2 r}\Bigr] +
\widehat{r}^i \widehat{r}^j \Bigl[ \frac{a''}{2} - \frac{a'}{2 r} +
\frac{b'}{2 r} - \frac{b}{r^2} \Bigr] \; . \label{ijcomp}
\end{eqnarray}
Comparing (\ref{00comp}) with (\ref{time}) implies,
\begin{equation}
b(r) = \frac{G M}{160 \pi^2 c^2} \times -\frac1{r^3} \; .
\end{equation}
Substituting this in (\ref{ijcomp}) and comparing with (\ref{space}) implies,
\begin{equation}
a(r) = \frac{G M}{160 \pi^2 c^2} \times \frac1{r^3} \; .
\end{equation}
Combining the classical and quantum corrections gives the following 
total results for the potentials,
\begin{eqnarray}
h_{00}(\vec{x}) & = & \frac{2 G M}{c^2 r} \Biggl\{ 1 +
\frac{\hbar G}{20 \pi c^3 r^2} + O\Bigl( \frac{\kappa^4}{r^4}\Bigr) \Biggr\} 
\; , \label{h00} \\
h_{ij}(\vec{x}) & = & \frac{2 G M}{c^2 r} \Biggl\{ 1 -
\frac{\hbar G}{20 \pi c^3 r^2} + O\Bigl( \frac{\kappa^4}{r^4}\Bigr) \Biggr\} 
\widehat{r}^i \widehat{r}^j \; . \label{hij}
\end{eqnarray}

\section{Discussion}

We have used the Schwinger-Keldysh formalism to derive manifestly
causal and real, linearized effective field equations (\ref{linoneeff}) 
for $h_{\mu\nu}(t,\vec{x})$ which include the one loop quantum effects 
of a massless, minimally coupled scalar. We used this equation to solve 
for the quantum corrections to the potentials (\ref{h00}-\ref{hij})
associated with a stationary point mass. Similar results have been obtained 
for conformally coupled scalars, fermions and vectors \cite{MJD,DL}. The
massless, minimally coupled scalar result does not seem to have been 
previously computed, but the real power of our analysis is the field 
equation (\ref{linoneeff}), which could be used even for a time dependent 
source. Our results (\ref{h00}-\ref{hij}) will also serve as an important 
correspondence limit for an on-going computation of the quantum corrected
potentials on a locally de Sitter background.

An important caveat to this analysis is that the full linearized effective 
field equations must include quantum corrections to the stress-energy of 
the source as well as corrections to the graviton kinetic operator. For
our problem, the scalar makes no contribution to the stress-energy of the
point mass which was our source. However, if we had considered quantum
gravitons it would have been necessary to correct the stress-tensor,
and some account would also have to be taken of the fact that the potentials
would be measured by coupling through a corrected vertex. These complications
are among the reasons why it required so much more work to derive reliable 
results for gravitons \cite{AFR,JFD,MK,HL,ABS,KK,gravpot}.

Finally, it is worth noting that the massless, minimally coupled scalar 
stands in isolation to its conformally coupled cousin. The self-energies of
both scalars take the same form,
\begin{eqnarray}
{\rm Minimally\ Coupled} & \!\!\!\! \Longrightarrow \!\!\!\! & 
\Bigl[ \mbox{}^{\mu\nu} \Sigma^{\rho\sigma}_{\scriptscriptstyle \pm \pm}
\Bigr] = (\pm) (\pm) \frac{i \kappa^2 {\rm D}^{\mu\nu\rho\sigma} 
\partial^2}{5120 \pi^4} \Biggl[\frac{\ln(\mu^2 \Delta x^2_{\scriptscriptstyle 
\pm \pm})}{\Delta x^2_{\scriptscriptstyle \pm \pm}} \Biggr] , \quad \\
{\rm Conformally\ Coupled} & \!\!\!\! \Longrightarrow \!\!\!\! & 
\Bigl[ \mbox{}^{\mu\nu} \widetilde{\Sigma}^{\rho\sigma}_{\scriptscriptstyle 
\pm \pm} \Bigr] = (\pm) (\pm) \frac{i \kappa^2 \widetilde{\rm D}^{\mu\nu\rho
\sigma} \partial^2}{5120 \pi^4} \Biggl[\frac{\ln(\mu^2 \Delta x^2_{
\scriptscriptstyle \pm \pm})}{\Delta x^2_{\scriptscriptstyle \pm \pm}} \Biggr] 
. \quad \label{conformal}
\end{eqnarray}
However, the 4th-order operator ${\rm D}^{\mu\nu\rho\sigma}$ has a nonzero 
trace whereas $\widetilde{\rm D}^{\mu\nu\rho\sigma}$ is traceless,
\begin{eqnarray}
{\rm D}^{\mu\nu\rho\sigma} & = & \Pi^{\mu\nu} \Pi^{\rho\sigma} + \frac13
\Pi^{\mu (\rho} \Pi^{\sigma) \nu} \; , \\
\widetilde{\rm D}^{\mu\nu\rho\sigma} & = & -\frac19 \Pi^{\mu\nu} 
\Pi^{\rho\sigma} + \frac13 \Pi^{\mu (\rho} \Pi^{\sigma) \nu} \; ,
\end{eqnarray}
where $\Pi^{\mu\nu} \equiv \eta^{\mu\nu} \partial^2 - \partial^{\mu}
\partial^{\nu}$. The graviton self-energy from a massless, 2-component 
fermion is three times that of the massless, conformally coupled scalar 
(\ref{conformal}), and the result for a massless vector is 12 times 
(\ref{conformal}), so the massless, minimally coupled scalar is the 
exceptional case. This distinction becomes even greater in de Sitter 
background on which massless, minimally coupled scalars experience
explosive particle production whereas the various massless, conformally
invariant particles do not. (For a simple discussion, see section III
of \cite{PW2}.)

\vskip .5cm

\centerline{\bf Acknowledgements}

It is a pleasure to acknowledge correspondence on this subject with
S. Deser and J. F. Donoghue. This work was partially supported by NSF 
grants PHY-0653085 and PHY-0855021 and by the Institute for Fundamental 
Theory at the University of Florida.

\end{document}